\documentstyle[12pt,aaspp4]{article}
 
\newcommand {\hI} {\ion{H}{1}\,\,}
\newcommand {\hII} {\ion{H}{2}\,\,}
\newcommand {\ha} {H$\alpha$\,\,}
\newcommand {\kms} {\,km\,s$^{-1}$\,}
\newcommand {\M} {\mbox{${\cal M}$}}

\newcommand {\msol} {\M$_\odot$\,}

\newcommand {\mlb} {(\M/L$_B$)$_\star$\,\,}

\newcommand {\mlsol}{\mbox{${\cal M}_\odot$/L$_{\odot}$}}
\newcommand {\mhIhe} {\M$_{HI+He}$\,}
\newcommand {\mldyn} {(\M/L$_B$)$_{dyn}$\,\,}
 
\begin{document}
 
\title{Accurate Parameters of the Mass Distribution in Spiral Galaxies: \\
1. Fabry--Perot Observations of NGC 5585}
 
\author{S\'ebastien Blais--Ouellette}
\affil{D\'epartement de physique and Observatoire du mont M\'egantic,
Universit\'e de Montr\'eal, C.P. 6128, Succ. centre ville,
Montr\'eal, Qu\'ebec, Canada. H3C 3J7 and\\
IGRAP, Observatoire de Marseille,
2 Place Le Verrier, F--13248  Marseille Cedex 04, France\\
e--mail: blaisous@astro.umontreal.ca}
\author{Claude Carignan\altaffilmark{1}} 
\affil{D\'epartement de physique and Observatoire du mont M\'egantic,
Universit\'e de Montr\'eal, C.P. 6128, Succ. centre ville,
Montr\'eal, Qu\'ebec, Canada. H3C 3J7\\
e--mail: carignan@astro.umontreal.ca}

\author{Philippe Amram}
\affil{IGRAP, Observatoire de Marseille,
2 Place Le Verrier, F--13248  Marseille Cedex 04, France\\
e--mail: amram@observatoire.cnrs-mrs.fr}
 
\author{St\'ephanie C\^ot\'e\altaffilmark{1}}
\affil{Dominion Astrophysical Observatory,
Herzberg Institute of Astrophysics,
National Research Council of Canada,
5071 West Saanich Rd., Victoria, BC, Canada. V8X~4M6\\
e--mail: Stephanie.Cote@hia.nrc.ca}

\altaffiltext{1}{Visiting Astronomers, Canada--France--Hawaii Telescope,
operated by the National Research Council of Canada, the Centre
National de la Recherche Scientifique de France, and the
University of Hawaii.}
 
\begin{abstract}

Using the example of the Sd galaxy NGC 5585, it is shown that high
resolution 2--D \hII kinematical data are necessary to determine
accurately the parameters of the mass (luminous \& dark) distribution in
spirals.  New CFHT Fabry--Perot \ha observations are combined with low
resolution (20~\arcsec) Westerbork \hI data to study its mass
distribution.  Using the combined rotation curve and best fit models, it
can be seen that \mlb of the luminous disk goes from 0.3 using only the
\hI rotation curve, to 0.8 using both the optical and the radio
data. This reduces the dark--to--luminous mass ratio in NGC 5585 by
$\sim 30$\% through increasing the dark matter halo core radius by
nearly the same amount.  This shows the importance of the inner, rising
part of the rotation curve for the accurate determination of the
parameters of the global mass (luminous \& dark) distribution and
suggests that such a fine tuning of the rotation velocities using high
resolution 2--D \hII kinematics is necessary to look at correlations
between the parameters of the dark matter component and other properties
of galaxies.

\end{abstract}
 
 
\section{INTRODUCTION}

In the last 25 years, a large number of rotation curves were
derived for spiral (Sp) and dwarf irregular (dIrr) galaxies from 2--D
\hI kinematics obtained with synthesis instruments such as the
Westerbork (WSRT) array, the Very Large Array (VLA), and the Australia
Telescope (AT) (for a good review of the first 20 years, see
e.g. Ashman 1992). In many galaxies, especially late--type spirals and
dwarf irregulars, the \hI extends much further out than the optical and thus than
the \hII emission.  An argument often used is that, since the \hI
rotation curve probes the gravitational potential in the dark matter
dominated region, it is best suited to derive the parameters of
the mass distribution and especially of the dark matter halo. However, as will
be shown, the parameters of the mass models (and especially of the dark matter
distribution) are very sensitive not only to the flat part of the
rotation curve (best probed by the \hI observations) but also to the
rising inner part, which can by derived with greater precision using
2--D \ha observations (see e.g. Amram et al. 1992, 1994, 1995,
1996). This is also well illustrated by Swaters (1999) who clearly
shows the impact of varying the position of the first few velocity
points (within the uncertainties due to beam smearing) on the
parameters of the mass models even in the dark matter dominated
dwarfs.

What is now regarded as the classical method to study the mass
distribution (van Albada et al. 1985, Carignan \& Freeman 1985) is
illustrated in Fig.\ref{fig:HI}a, which shows the analysis of the mass
distribution of NGC 5585 using its \hI rotation curve (C\^ot\'e,
Carignan, \& Sancisi 1991).  See also Begeman 1987, Broeils 1992 and
C\^ot\'e 1995 for many more examples. First, the rotation curve is
obtained by fitting a ``tilted--ring'' model to the \hI velocity field
in order to represent the warp of the \hI disk, which is almost always
present. The accuracy of the model representation is then checked by
looking at the residual (data $-$ model) map (Warner 1973, Sancisi \&
Allen 1979). Then the luminosity profile in the reddest band
available to probe the mass dominant population is transformed into a
mass distribution for the stellar disk, assuming a constant value of
\mlb (Casertano 1983, Carignan 1985). For the contribution of the
gaseous component, the \hI radial profile scaled by 1.33 is used to
account for He. The difference between the observed rotation curve and the
computed contribution to the curve of the luminous (stars \& gas)
component is thus the contribution of the dark component, which can be
represented by an isothermal halo (Carignan 1985) or some other
functional form (e.g. Lake \& Feinswog 1989). The model of
Fig.\ref{fig:HI}a allows us to study the dark--to--luminous mass ratio
as a function of radius, as shown in Fig.\ref{fig:HI}b.  Naturally,
this is for standard gravity. Some alternative models, such as MOND,
have also been explored (Milgrom 1983, Sanders 1996, McGaugh \& de Blok 1998).

The example of NGC 5585 shows the importance of an accurate
determination of the rising part of the rotation curve, since this is
the part that mainly constrains the values of two of the three free
parameters of the mass model; namely, the mass--to--light ratio of the
luminous stellar disk \mlb and the core radius r$_c$. The third
parameter, the one dimensional velocity dispersion $\sigma$ of the
dark isothermal halo is mainly constrained by the outer part of the
rotation curve.  The \hI observations, often optimized for maximum
sensitivity in the outer parts, have in most of the published studies
a resolution of only 20--45 \arcsec\, (higher resolution is naturally
possible by adding longer baselines when there is sufficient
\hI flux).  Attempts have been made to correct for the effect of
``beam smearing'', which can be very important in the inner parts
because of the strong velocity gradient (sometimes combined with a
strong radial distribution gradient) across the large \hI beam. This
is examined using as an example the Sc galaxy NGC 3198.

Another point that needs to be stressed is that full 2--D
\hII kinematical data are necessary for this work and that 1--D long--slit
spectroscopy is not sufficient. This is due to the fact that the
photometric parameters (we are mainly concerned with the position
angle PA and photometric center in this case) used to position the
slits on the galaxies can sometime be quite different from the
kinematical parameters. Naturally, if the slit is positioned with a
slightly wrong PA, the velocities will necessarily be underestimated.
This is well illustrated for the case of the rotation curves of
galaxies in clusters (Whitmore, Forbes \& Rubin 1988 for the 1--D
long-slit, and Amram et al. 1996 for 2--D Fabry--Perot).

The importance of the rising part of the rotation curve on the
parameters for both the luminous and dark matter distributions is
illustrated by two examples in section 2. Section 3 describes the new
CFHT Fabry--Perot (FP) observations and data reduction of the NGC 5585
data. The \hII kinematics and the optical rotation curve are discussed
in Section 4, while the mass models and the parameters of the mass
distribution are given in Section 5. Finally, Section 6 gives a
summary of the results and draws general conclusions from this study.

\section{IMPORTANCE OF THE RISING PART OF THE ROTATION CURVE ON THE
PARAMETERS OF THE MASS (LUMINOUS \& DARK) DISTRIBUTION}

It has always been thought that the problem of ``beam smearing'' was
important mainly in early--type spirals, where the strong gradient due to the
presence of the bulge was attenuated in low resolution \hI data and
where it was obvious that higher resolution data were necessary to see
the true kinematics resulting from the centrally concentrated luminous
mass distribution. In what follows, it will be shown that, while the
effect of beam smearing in late--type spirals may be less
dramatic, it can nevertheless have a significant impact on the derived
parameters of both the luminous and the dark mass distributions.

\subsection{The Case of NGC 5585}

To show the importance of the first few points of the rotation curve
in a galaxy such as NGC 5585, a model was constructed giving no weight
to the first two points of the \hI curve (Fig.\ref{fig:HI-2first}a).
This model mimics a difference of less than 10 \arcsec~ with the real
position of the first two points, a very plausible effect of the large radio
beams. In this model, the \mlb of the stellar disk goes from 0.3
(Fig.\ref{fig:HI}a) to 1.0 (Fig.\ref{fig:HI-2first}a), with the result
that the mass of the stellar disk goes from $\sim$20\% of the gaseous
disk to a comparable mass. More importantly is that the dark matter
halo is less centrally concentrated with a dark--to--luminous mass
ratio going from 9.5 (Fig.\ref{fig:HI}b) to 6.3
(Fig.\ref{fig:HI-2first}b) at the last measured point of the rotation
curve.  This is a difference of more than 30\% in the
dark--to--luminous mass ratio for a difference of less than 10 \arcsec
\, in the position of the first two points of the curve.  As
illustrated in Fig.\ref{fig:HI}b \& Fig.\ref{fig:HI-2first}b, the
global distribution of the dark component is also totally different.
This is why we think that the ideal rotation curve to study the mass
distribution in galaxies should combine the high resolution of \ha FP
observations in the inner parts to the high sensitivity of the low
resolution \hI observations in the outer parts.

\subsection{The Case of NGC 3198}

Begeman (1989) published a Westerbork \hI rotation curve of NGC 3198,
where he attempted to correct for the effect of
beam--smearing. Theoretically, one should be able to calculate this
effect by convolving the rapidly dropping HI density profile and the
rising rotation curve inside the width of the beam.  In the inner
parts, his rotation velocities are systematically larger (up to 26
\kms at 30 \arcsec) than the values derived in a previous \hI study by
Bosma (1981). If the corrections are accurate, one would expect that
there should be very little gain in using high resolution \ha data.
Fig.\ref{fig:3198Beg} and Table~\ref{tab:mod3198} show the best--fit
model using the beam--smearing corrected \hI data. It can be seen that
for r~$<$~3~kpc and r~$>$~15~kpc, the model gives a good
representation of the data. However, around 4 kpc, the model velocity
is larger by $\sim 10$
\kms compared to the measured velocity.

A best-fit model (Fig.\ref{fig:3198Beg+Cor}) was obtained by combining
Begeman's \hI data with the FP \ha kinematical data of Corradi et
al. (1991).  We see that while the agreement between the two sets of
data appears good over all, the optical velocities are somewhat
smaller in the steep rising part of the rotation curve.  As can be seen in
Table~\ref{tab:mod3198}, the dark--to--luminous mass ratio at the last
measured point has changed very little between the two models (2.9
$\rightarrow$ 3.0), but the shape of the halo has changed
substantially, becoming more centrally concentrated with r$_c$ going
from 17.2 to 11.7 kpc, again a change of more than 30\%. The
apparently small difference in velocity ($\sim 5$ \kms) results in an
increase of the dark halo central density $\rho_0$ by nearly a factor
of 2 (0.004 $\rightarrow$ 0.008).  This suggests that Begeman (1989)
may have overestimated his beam--smearing corrections.

It is instructive also to compare this result with the earlier Bosma
data, which were not corrected for beam--smearing, as is the case for
most \hI data. NGC 3198 is an Sc galaxy, in which the velocity
gradient is much smaller than in Sa or Sb galaxies and one would have
thought that the effect of beam--smearing should not be that
dramatic. Fig.\ref{fig:3198Bos} shows the best fit model using that
data set. We see that the mass distribution is completely different,
with a much smaller disk and a dark halo that dominates completely for
r~$\ge 1$~kpc. The result is that, with differences $\le 10$ \kms for
$0 \le {\rm r} \le 6$~kpc, the dark component has nearly 10 times
higher central density, which results in an increase of the
dark--to--luminous mass ratio from $\sim 1$ to $\sim 4$.

Many more examples could be discussed, but we think that the examples
above show clearly that high resolution \ha data are necessary to
compute accurately the parameters of both the luminous and dark mass
distributions.

\section{FABRY--PEROT OBSERVATIONS \& REDUCTION}

Table~\ref{tab:opt5585} gives the optical parameters of NGC 5585 and
Table~\ref{tab:fp5585} lists the complete observing parameters. The FP
observations of the \ha emission line were obtained in February 1994
at the Canada--France--Hawaii Telescope (CFHT). The FP etalon (CFHT1)
was installed in the CFHT's Multi--Object Spectrograph (MOS).  A
narrow--band filter ($\Delta \lambda$ = 10\,\AA), centered at
$\lambda_0$ = 6570\,\AA\, (nearly at the systemic velocity of NGC
5585, V$_{sys} \approx 305$ \kms), was placed in front of the etalon.
The available field with no vignetting was $\approx$ 8.5\arcmin
$\times$ 8.5\arcmin, with 0.34\arcsec\, pix $^{-1}$. The free spectral
range of 5.66\,\AA\, (258 \kms) was scanned in 27 (+1 overlapping)
channels, giving a sampling of 0.2\,\AA\, (9.2 \kms) per
channel. Eight minutes integration was spent at each channel position.

\subsection{Data analysis}

Following normal de--biasing and flat--fielding with standard
IRAF procedures, a robust 3-D cosmic--ray removal routine, 
that tracks cosmic rays by spatial (pixel--to--pixel) and spectral
(frame--to--frame) analysis, was applied.

Since FP systems have multiple optical surfaces, some defocalised
ghost reflections can be present (Bland-Hawthorn 1995), especially
since the etalon was not tilted.
To get rid of these reflections we composed a "ghost image" by
using the ghost reflection of a bright star in the field (Figure
\ref{fig:reflex}) and numerically simulating a similar but scaled 
reflection for every pixel in the field. This image was then
subtracted from the original. This procedure removes very efficiently
all the reflected continuum and adequately but not perfectly
($\sim$80\%) the monochromatic emission.

The presence of strong night sky lines combined with photometric
variations (transparency, seeing) from one exposure to another led us
to proceed to a first background subtraction on each of the 27
non-redundant frames (now assembled in a 3-D cube).  This background
includes continuous, diffuse light and monochromatic emission from
atmospheric OH radicals and from geocoronal H$\alpha$. All these
background vary both spatially and temporally. Using the radial symmetry of
the FP, the sky was evaluated by azimuthally summing rings of constant
phase where the galaxy signal had been masked. The computed background
was then removed in each ring.

A neon calibration lamp ($\lambda$6598.95 \AA) was used to fix the
zero point at each pixel. To be totally device independent, the
theoretical position of a sky emission line was then used to
fine-tuned the phase (wavelength origin) at each pixel in order to get
a particular wavelength on an exact x-y plane. Due to limited free
spectral range, this telluric line is a composite of geocoronal
\ha ($\lambda$6562.74 or 517 \kms) and an OH line ($\lambda$6568.78 or
532 \kms). Since there is no way to determine the relative
contribution of each line, we are left with some uncertainties on the
systemic velocity of the galaxy, but this does not affect the relative
velocities and the rotation curve.

In order to get sufficient signal--to--noise throughout the image, two
different Gaussian smoothings ($\sigma$=2.5 and 3.5 pixels) were performed
on the cube using the ADHOC package (Boulesteix 1993). Velocity maps were then 
obtained using the intensity weighted means of the H$\alpha$ peaks to 
determine the radial velocity for each pixel. A final variable
resolution velocity map was constructed (Figure \ref{fig:vit+mono}) using
higher resolution for regions with originally higher signal-to-noise.

\section{HII KINEMATICS \& OPTICAL ROTATION CURVE}

The rotation curve has been obtained from the velocity field following
two different methods. The first estimate was made using the task
ROCUR (Begeman 1987, Cot\'e et al. 1991) in the AIPS package, where
annuli in the plane of the galaxy (ellipses in the plane of sky) are
fitted to the velocity field, minimizing the dispersion inside each
ring. In this way, the center, systemic velocity, position angle and
inclination are evaluated.  Secondly, the ADHOC package was used to
fine--tune these parameters by direct visualization and comparison
with a residual velocity field.  The optical rotation curve at
5\arcsec~resolution is given in Table~\ref{tab:rc5585} and
Figure~\ref{fig:rc_Ha}. Note that there are two common ways to
represent the errors on a rotation curve: the error on the mean
($\sigma/\sqrt{N})$) and the velocity difference of the receding and
approaching side weighted by the number of points on each side, a
method often used for HI rotation curve.  To be conservative, we took
the maximum of the two values.

At intermediate radii, the approaching side of the galaxy is still
affected by residual sky emission. This is caused by the lack of
regions with pure sky signal in the most central rings, making the
measurement of the sky emission lines less accurate and only partially
subtracted. The final effect here is to lower the rotation velocities
between 2 and 4 kpc in radius. As we will see, this is in the region
where it is possible to rely with confidence on the \hI data, because of
the shallower fall of \hI density and the slower rise of the rotation
curve that make beam smearing negligible.

\section{MASS MODELS AND PARAMETERS OF THE MASS DISTRIBUTION} 

The models used are described in Carignan (1985). However, instead of
being ``maximum disk'' models, they are ``best--fit'' models. A $\chi^2$
minimization technique is used in the three--parameter space of the
model. Namely, those parameters are: \mlb of the stellar disk,
the core radius r$_c$ and the one--dimensional velocity dispersion $\sigma$
of the dark isothermal halo. Alternatively, one can use the central
density $\rho_0=9\sigma^2/4\pi G r_c^2$. The surface photometry and
the \hI kinematics are from C\^ot\'e, Carignan, \& Sancisi 1991.

\subsection{Mass Model from the \ha Rotation Curve}

The best--fit mass model for the \ha rotation curve at 5\arcsec\,
resolution is shown in Fig.\ref{fig:Ha}. It can be seen that there is
a clear sign of the disk mass in the rotation curve, which is well
fitted. In fact, the best--fit model is essentially a maximum disk
model. The mass--to--light ratio of the stellar disk goes from
0.3 using the \hI data to 1.0 using the \ha data, which causes the
halo to become less centrally concentrated. For the dark halo, the
parameters are r$_c$ = 4.1 kpc, $\rho_0$ = 0.023 \msol pc$^{-3}$ and
$\sigma$ = 49.1 \kms, which represent a decrease of $\rho_0$ of more
than 50\%. Interestingly, the \ha rotation curve provides a much
better fit to the MOND model ($a_0 = 1.2 \times 10^{-8} cm s^{-2}$,
M$^{\ast}$/L$_B$ = 0.5) than the \hI curve alone (see figure 1 of
Sanders 1996).  However, the little kink seen at radius $\simeq$1 kpc
could indicate the transition between the disk dominated region and
the halo dominated region, which would exclude alternative
gravitational theories based on luminous matter only. This feature
could also be the dynamical signature of an inner bar, but the 2-D
velocity field does not show evidence of non-circular motion.

It is interesting to look at the shape of the different components as
a function of radius for this \ha rotation curve, derived out to
$\sim$1.3 R$_{25}$ (herein defined as RC3 D$_{25}$/2) or $\simeq
3.3$\arcmin $\simeq 6.0$ kpc. In massive spirals, the stellar disk
usually dominates the mass distribution for r $<$ R$_{25}$. Typical
\M$_{dark}$/\M$_{lum}$ are between 0.5 to 1.0 at that radius. This is
certainly not the case here with
\M$_{dark}$/\M$_{lum}$ $\simeq 4.0$ at the last measured point of the
rotation curve. Moreover, at the last point, there is almost as much
luminous mass in gas as in stars.  So, for a dwarf spiral such as NGC
5585, the mass distribution is much more reminiscent of what is seen
in dIrr (e.g. DDO 154: Carignan \& Freeman 1988, Carignan \& Beaulieu
1989; DDO 170: Lake, Shommer, \& van Gorkom 1990) than in massive Sp
galaxies ( e.g. NGC 6946: Carignan et al. 1990; NGC 3198: van Albada
et al. 1985). Other late--type Sp such as IC 2574 (Martimbeau et
al. 1994) and NGC 3109 (Jobin \& Carignan 1990), both of type Sm, also
have a strong contribution from dark matter even in the inner parts
but show solid-body \hI rotation curves.

\subsection{Mass Model from the Combined \hI \& \ha Rotation Curve}

Table~\ref{tab:mod5585} gives the parameters of the mass models
constructed using only the \hI rotation curve, only the \ha curve, and
the combined \hI \& \ha curve.  For our adopted mass model of NGC
5585, we combine the high resolution of the \ha data in the inner
parts with the high sensitivity of the \hI data in the outer
parts. Since we are making a best--fit model, one has to
understand that, because of the higher resolution, there are more \hII
data points than \hI data points. This means that the optical data
would tend to have a higher weight than the radio data. Since optical
velocities are derived from high S/N data out to a radius of
120\arcsec\, and since Fig. 9 of C\^ot\'e, Carignan \& Sancisi (1991)
shows that this is the region where the \hI parameters are not well
defined, we decided to use for the final model the \ha data for
r~$<$~120\arcsec\, and the \hI data for r~$>$~120\arcsec.

This adopted model is shown in Fig.\ref{fig:HI+Ha}. The
parameters of the model are: \mlb = 1.0, r$_c$ = 4.5 kpc, $\rho_0$ =
0.024 \msol pc$^{-3}$ and $\sigma$ = 53.6 \kms. As expected,
$\sigma$ is very similar in the combined \hI \& \ha curve as in the
\hI rotation curve. This is the case because this parameter is a
measure of the maximum amplitude of the rotation curve, which is
mainly defined by the \hI data in the outer parts. However, the two
other parameters \mlb for the stellar disk and $\rho_0$ of the dark
halo (which are coupled) have nearly the same values as those derived
with the \ha curve. Again, this is because
\mlb of the luminous stellar disk, and hence the scaling parameter 
of the dark halo r$_c$, is mainly constrained by the \hII data in the
inner parts. Interestingly, because this newly derived central density
is significantly lower, this means that this late-type galaxy's dark
halo is even less concentrated; therefore this exacerbates the
discrepancy between observed rotation curves and those predicted by
standard CDM halo simulations , which are already too concentrated for
late-type and dwarf galaxies (see, e.g., Navarro 1996 but also
Kravtsov et al. 1998).

\section{SUMMARY AND CONCLUSIONS}

The importance of an accurate determination of the rising part of
a rotation curve using full 2--D high resolution FP
observations is well illustrated by the example of NGC 5585. 
The principal conclusions follow.

1. The parameters of the mass distribution of both the dark and the
luminous components are very sensitive to the rising part of the rotation curve
(the first few velocity points) not only in early-type spirals, where
the velocity gradient is large in the inner parts, but also in
late-type spirals, which have a much shallower gradient. The
sensitivity is especially important when the contributions of dark and
luminous matter are comparable.

2. With the example of NGC 3198, it is shown that it is very difficult to
correct theoretically for the beam--smearing effect seen in radio data.

3. Full 3-D spectroscopy, obtained with Fabry-Perot spectroscopy, is
to be preferred to long-slit spectroscopy in order to derive properly
the orientation parameters (namely, the rotation center and the
position angle) and hence not underestimate the rotational velocities.

4. Combining new \ha CFHT FP data with Westerbork HI data reduced the
ratio \M$_{dark}$/\M$_{lum}$ by $\simeq$ 30\% via a decrease of the
central density by nearly a factor of 3 for the late-type spiral NGC
5585. If such large errors are common, one could imagine that it could
mask any physical correlation between the parameters of the dark and
the luminous matter.

5. Finally, the optimal rotation curve is clearly a combination of 2--D high
resolution spectroscopy for the inner part of spiral galaxies and high
sensitivity radio observations for the outer regions. 

\acknowledgments
 
We would like to thank the staff of the CFHT for their support during
the FP data acquisition and Daniel Durand from DAO who helped with
data acquisition. We also warmly thank Jacques Boulesteix for
fuitfull discussion on Fabry-Perot reduction and Anthony F.J. Moffat
for useful comments. CC acknowledges grants from NSERC (Canada) and
FCAR (Qu\'ebec).

\clearpage

\begin{deluxetable}{l c c c}
\tablenum{1}
\tablecaption{Parameters of the mass models of NGC 3198.\label{tab:mod3198}}
\tablehead{
\colhead{Parameter}
&\colhead{\hI\tablenotemark{a} RC}
&\colhead{Combined \hI\tablenotemark{a} \& \ha RC}
&\colhead{\hI\tablenotemark{b} RC}
}

\startdata
{\it Luminous disk component}:						  \nl
\mlb \hfill	(\mlsol) 	&9.4 $\pm 0.2$  	&8.5 $\pm 0.3$ 
&2.8 $\pm 0.5$ \nl
\M$_\star$ \hfill(\msol)	&$3.2 \times 10^{10}$	&$2.9 \times 10^{10}$
&$9.6 \times 10^9$ \nl
\mhIhe \hfill	(\msol)	 	&$6.5 \times 10^9$	&$6.5 \times 10^9$
&$6.5 \times 10^9$\nl
\nl
{\it Dark halo component}:						  \nl
r$_c$ \hfill	(kpc)		&17.2 $\pm 1.0$		&11.7 $\pm 1.0$	  
&3.9 $\pm 0.1$ \nl
$\sigma$ \hfill	(\kms)		&85.6 $\pm 2.0$		&79.0 $\pm 1.5$	  
&83.4 $\pm 1.0$ \nl
$\rho_0$ \hfill	(\msol pc$^{-3}$)&0.004			&0.008		  
&0.076 \nl
\nl
{\it At R$_{HO}$ r $\simeq 13$ kpc}:						  \nl
$\rho_{halo}$	\hfill (\msol pc$^{-3}$)&0.002		&0.002		  
&0.002 \nl
\M$_{dark+lum}$ \hfill (\msol) &$6.2 \times 10^{10}$ &$6.6 \times 10^{10}$
&$6.6 \times 10^{10}$ \nl
\mldyn				&18			&19		  
&19.5 \nl
\M$_{dark}$/\M$_{lum}$		&0.76			&1.1		  
&4.3 \nl
\nl
{\it At the last measured point r $\simeq 29$ kpc}:                         \nl
$\rho_{halo}$ \hfill (\msol pc$^{-3}$)&0.0005    &0.0004  &                 \nl
\M$_{dark+lum}$ \hfill (\msol) &$1.5 \times 10^{11}$ &$1.4 \times 10^{11}$& \nl
\mldyn                        &44                      &41 	&           \nl
\M$_{dark}$/\M$_{lum}$         &2.9                     &3.0 	&            \nl

\nl
\tablenotetext{a}{Begeman 1989}
\tablenotetext{b}{Bosma 1981}
\enddata
\end{deluxetable}

\clearpage

\begin{deluxetable}{lr}
\tablenum{2}
\tablewidth{25pc}
\tablecaption{Optical  parameters of NGC 5585.\label{tab:opt5585}}
\tablehead{}

\startdata
Morphological Type\tablenotemark{a}		&SABd			\nl
RA (J2000.0)			  &14$^{\rm h}$ 19$^{\rm m}$ 48\fs 1	\nl
Dec (J2000.0)		 		&56\arcdeg 43\arcmin 44\arcsec  \nl
l						&214 \fdg 95		\nl
b						&56 \fdg 73		\nl
Adopted distance (Mpc)\tablenotemark{b}		&6.2 			\nl
						&(1\arcmin\ $\simeq 1.8$ kpc)\nl
Mean axis ratio, q = b/a\tablenotemark{c}		&0.61 $\pm 0.01$\nl
Inclination($\rm q_0$ = 0.12), i\tablenotemark{c}&53 \arcdeg $\pm 1$ \arcdeg \nl
Isophotal major diameter, D$_{25}$\tablenotemark{c}	&5.27 \arcmin	\nl
Major axis PA\tablenotemark{c}			&99 \arcdeg $\pm 1$\arcdeg\nl
Exponential scale length (kpc)\tablenotemark{c} &1.4			\nl
Holmberg radius, R$_{\rm HO}$\tablenotemark{c}  & 3.62 \arcmin		\nl
Absolute magnitude, M$_B$\tablenotemark{c}	&--17.5 		\nl
Total luminosity, L$_B$			       &$1.5 \times 10^9$ L$_{\odot}$\nl
Helio. radial velocity (\kms)\tablenotemark{a}	&305 $\pm 3$		\nl
\tablenotetext{a}{de Vaucouleurs et al. (1991).}
\tablenotetext{b}{$\rm H_0$ = 75 \kms Mpc$^{-1}$.}
\tablenotetext{c}{C\^ot\'e, Carignan, \& Sancisi (1991).}
\enddata
\end{deluxetable}

\clearpage

\begin{deluxetable}{lr}
\tablenum{3}
\tablewidth{25pc}
\tablecaption{Parameters of the Fabry--Perot observations.\label{tab:fp5585}}
\tablehead{}

\startdata
Date of observations        &           February 20, 1994	    	\nl
Telescope                   &                        3.6\,m CFHT  	\nl
Instrumentation: \hspace{8cm}& 						\nl
~~~~~Focal plane instrument &		MOSFP				\nl
~~~~~CCD detector           &            2048\,$\times$\,2048 Loral3,
                                         $\sigma$ = 8\,e$^{-1}$   	\nl
~~~~~Filter                 &            $\lambda_0$ = 6570\,\AA,
                                                 $\Delta \lambda$ = 10\,\AA \nl
~~~~~Fabry--Perot etalon    &           Scanning QW1162 (CFHT1) 	\nl
~~~~~Interference order     &            1155 @ 
			$\lambda_{\scriptscriptstyle N\!E\!O\!N}$ 	\nl
~~~~~Mean Finesse in the field &  	 12				\nl
~~~~~Calibration lamp       &        Neon ($\lambda$ = 6598.95\,\AA)	\nl
Duration                    &						\nl
~~~~~Per channel		    &            8\,min/channel 		\nl
~~~~~Total                  &		 3\,h~45\,min			\nl
Spatial Parameters:  	    &						\nl
~~~~~Field size             &            8.5$'$\,$\times$\,8.5$'$   	\nl
~~~~~Pixel scale               &            0.34$''$\,pix$^{-1}$      	\nl
Spectral Parameters: 	    &						\nl 
~~~~~Number of channels     &                             27   		\nl
~~~~~Free spectral range    &     5.66\,\AA\ (258\,km\,s$^{-1}$) 	\nl
~~~~~Sampling    	    &     0.2\,\AA\ (9.2\,km\,s$^{-1}$)/channel \nl
\enddata
\end{deluxetable}

\clearpage

\begin{deluxetable}{ccccccccc}
\tablenum{4}
\tablewidth{35pc}
\tablecaption{Optical  rotation curve at 5\arcsec\, resolution\tablenotemark{1}.
\label{tab:rc5585}}
\tablehead{
\colhead{Radius}&\colhead{N$_{app}$}&\colhead{V$_{app}$}&\colhead{N$_{rec}$}
&\colhead{V$_{rec}$}&\colhead{V$_c$}\\
\colhead{(arcsec)}&&\colhead{\kms}&&\colhead{\kms}&\colhead{\kms}}
\startdata
  2.5  &  27   & 10 $\pm$ 2 &  20   &  9  $\pm$ 2  & 11  $\pm$ 2  \nl
  7.5  &  70   & 26 $\pm$ 1 &  59   & 25  $\pm$ 2  & 26  $\pm$ 1  \nl
 12.5  & 104   & 33 $\pm$ 1 & 100   & 34  $\pm$ 1  & 33  $\pm$ 1  \nl
 17.5  & 160   & 31 $\pm$ 1 &  99   & 33  $\pm$ 1  & 32  $\pm$ 1  \nl
 22.5  & 198   & 33 $\pm$ 1 &  86   & 34  $\pm$ 1  & 33  $\pm$ 1  \nl
 27.5  & 201   & 36 $\pm$ 1 & 100   & 32  $\pm$ 1  & 35  $\pm$ 2  \nl
 32.5  & 195   & 40 $\pm$ 1 & 137   & 34  $\pm$ 1  & 37  $\pm$ 3  \nl
 37.5  & 217   & 44 $\pm$ 1 & 173   & 38  $\pm$ 1  & 41  $\pm$ 3  \nl
 42.5  & 211   & 44 $\pm$ 1 & 131   & 41  $\pm$ 1  & 43  $\pm$ 2  \nl
 47.5  & 206   & 43 $\pm$ 1 & 125   & 46  $\pm$ 1  & 44  $\pm$ 1  \nl
 52.5  & 194   & 46 $\pm$ 1 &  93   & 46  $\pm$ 1  & 46  $\pm$ 1  \nl
 57.5  & 178   & 47 $\pm$ 1 & 103   & 45  $\pm$ 1  & 46  $\pm$ 1  \nl
 62.5  & 193   & 46 $\pm$ 1 &  91   & 48  $\pm$ 1  & 46  $\pm$ 1  \nl
 67.5  & 225   & 51 $\pm$ 1 &  54   & 45  $\pm$ 2  & 50  $\pm$ 3  \nl
 72.5  & 270   & 57 $\pm$ 1 &  62   & 54  $\pm$ 2  & 56  $\pm$ 1  \nl
 77.5  & 267   & 57 $\pm$ 1 &  72   & 53  $\pm$ 2  & 56  $\pm$ 2  \nl
 82.5  & 285   & 61 $\pm$ 1 &  29   & 62  $\pm$ 2  & 61  $\pm$ 2  \nl
 87.5  & 265   & 64 $\pm$ 1 &   6   & 43  $\pm$ 8  & 64  $\pm$ 6  \nl
 92.5  & 288   & 66 $\pm$ 1 &  18   & 60  $\pm$ 4  & 66  $\pm$ 2  \nl
 97.5  & 196   & 68 $\pm$ 1 &  70   & 62  $\pm$ 2  & 66  $\pm$ 3  \nl
 102.5 &  86   & 71 $\pm$ 1 &  36   & 51  $\pm$ 4  & 67  $\pm$ 9  \nl
 107.5 & 131   & 73 $\pm$ 1 &  17   & 55  $\pm$ 7  & 72  $\pm$ 7  \nl
 112.5 & 105   & 72 $\pm$ 1 &   4   & 60  $\pm$ 2  & 72  $\pm$ 5  \nl
 117.5 &  89   & 72 $\pm$ 1 &  37   & 61  $\pm$ 2  & 69  $\pm$ 5  \nl
 122.5 &  86   & 74 $\pm$ 1 &  38   & 68  $\pm$ 3  & 73  $\pm$ 3  \nl
 127.5 & 121   & 76 $\pm$ 1 &  48   & 59  $\pm$ 3  & 73  $\pm$ 8  \nl
 132.5 & 179   & 73 $\pm$ 1 &  52   & 61  $\pm$ 1  & 70  $\pm$ 5  \nl
 137.5 & 170   & 88 $\pm$ 1 &  62   & 65  $\pm$ 1  & 82  $\pm$ 1  \nl
 142.5 & 160   & 87 $\pm$ 1 &  27   & 79  $\pm$ 1  & 86  $\pm$ 4  \nl
 147.5 & 124   & 82 $\pm$ 1 &  56   & 77  $\pm$ 1  & 80  $\pm$ 2  \nl
 152.5 &  72   & 84 $\pm$ 1 &  27   & 75  $\pm$ 2  & 81  $\pm$ 4  \nl
 157.5 &  24   & 85 $\pm$ 1 & 113   & 76  $\pm$ 1  & 77  $\pm$ 4  \nl
 162.5 &  44   & 82 $\pm$ 1 &  80   & 77  $\pm$ 1  & 79  $\pm$ 2  \nl
 167.5 &  29   & 83 $\pm$ 1 &  35   & 79  $\pm$ 2  & 81  $\pm$ 2  \nl
 172.5 &   1   & 86 $\pm$ 1 &  17   & 87  $\pm$ 2  & 87  $\pm$ 2  \nl
 177.5 &   5   & 78 $\pm$ 5 &  67   & 77  $\pm$ 2  & 77  $\pm$ 2  \nl
 182.5 &  12   & 90 $\pm$ 6 &  19   & 77  $\pm$ 3  & 82  $\pm$ 7  \nl
 187.5 &   1   & 88 $\pm$ 1 &  20   & 79  $\pm$ 2  & 80  $\pm$ 5  \nl
 192.5 &   0   &            &  30   & 80  $\pm$ 2  & 80  $\pm$ 2  \nl
 197.5 &   0   &            &  13   & 73  $\pm$ 4  & 73  $\pm$ 4  \nl

\tablenotetext{1}{derived with i = 52\arcdeg, PA = 43\arcdeg }
\enddata
\end{deluxetable}

\clearpage

\begin{deluxetable}{l c c c}
\tablenum{5}
\tablecaption{Parameters of the mass models of NGC 5585.\label{tab:mod5585}}
\tablehead{
\colhead{Parameter}&\colhead{\hI RC}&\colhead{\ha RC}
&\colhead{Combined \hI \& \ha RC}
}

\startdata
{\it Luminous disk component}:						  \nl
\mlb 	\hfill	(\mlsol) &0.3 $\pm 0.3$ \tablenotemark{a} &1.0 $\pm 0.1$ &1.0 $\pm 0.1$  \nl
\M$_\star$ \hfill (\msol)	&$3.3 \times 10^8$	&$1.1 \times 10^9$ &$9.9 \times 10^8$\nl
\mhIhe\hfill	(\msol)	 	&$1.4 \times 10^9$	&$1.4 \times 10^9$ &$1.4 \times 10^9$\nl
\nl
{\it Dark halo component}:						  \nl
r$_c$ \hfill	(kpc)		&2.8 $\pm 0.3$		&4.1 $\pm 0.4$ &4.3 $\pm 0.4$	  \nl
$\sigma$ \hfill	(\kms)		&52.9 $\pm 2.0$		&49.1 $\pm 2.0$ &53.6 $\pm 1.6$	  \nl
$\rho_0$ \hfill	(\msol pc$^{-3}$)&0.060		&0.023   &0.024		  \nl
\nl
{\it At R$_{HO}$ r = 6.5 kpc}:						  \nl
$\rho_{halo}$ \hfill (\msol pc$^{-3}$)&0.0035		& &0.0041		  \nl
\M$_{dark+lum}$ \hfill (\msol) &$1.2 \times 10^{10}$ & &$1.1 \times 10^{10}$\nl
\mldyn				&10.6			& &10.1		  \nl
\M$_{dark}$/\M$_{lum}$		&8.7			& &4.6		  \nl
\nl
{\it At the last measured point r = 9.6 kpc}:				  \nl
$\rho_{halo}$ \hfill (\msol pc$^{-3}$)&0.0013		& &0.0017		  \nl
\M$_{dark+lum}$ \hfill (\msol) &$1.7 \times 10^{10}$ & &$1.8 \times 10^{10}$\nl
\mldyn                        &15.7 			& &16.4		  \nl
\M$_{dark}$/\M$_{lum}$         &9.5		&	& 6.6		  \nl
\tablenotetext{a}{The difference in \mlb between this paper and 
C\^ot\'e, Carignan, \& Sancisi (1991) comes from using a different
Galactic extinction value, A$_B$ = 0.0 (RC3).}
\enddata
\end{deluxetable}

\clearpage

\figcaption[fig1.ps]
{{\bf a)} Best fit mass model for NGC 5585 using the \hI rotation curve. 
The model parameters are: \mlb = 0.3, $r_c$ 2.8 kpc and $\sigma$ = 53 \kms. \\
{\bf b)} Dark--to--luminous mass ratio as a function of radius.\label{fig:HI}}

\figcaption[fig2.ps]
{{\bf a)} Maximum disk mass model for NGC 5585, where the first two points of
the \hI rotation curve have been given zero weight. The model parameters are:
\mlb = 1.0, $r_c$ = 3.5 kpc and $\sigma$ = 52 \kms. \\
{\bf b)} Dark--to--luminous mass ratio as a function of radius.
\label{fig:HI-2first}}

\figcaption[fig3.ps]
{Best fit mass model for NGC 3198 using the \hI rotation curve (Begeman 1989),
corrected for beam--smearing.
The model parameters are: \mlb = 9.4, $r_c$ = 17.2 kpc and $\sigma$ = 85.6 \kms. 
\label{fig:3198Beg}}

\figcaption[fig4.ps]
{Best fit mass model for NGC 3198 using the \hI (filled circles) rotation 
curve (Begeman 1989) and the \ha (open circles) rotation curve (Corradi 
et al. 1991).
The model parameters are: \mlb = 8.5, $r_c$ = 11.7 kpc and $\sigma$ = 79.0 \kms. 
\label{fig:3198Beg+Cor}}

\figcaption[fig5.ps]
{Best fit mass model for NGC 3198 using the \hI rotation curve of Bosma (1981),
not corrected for beam--smearing.
The model parameters are: \mlb = 2.8, $r_c$ = 3.9 kpc and $\sigma$ = 83.4 \kms.
\label{fig:3198Bos}}

\figcaption[fig6.ps]
{{\bf a)} Real reflection of a star.
{\bf b)} Cut along the y axis of the real reflection. {\bf c)} Cut along the 
y axis of the simulated reflection. \label{fig:reflex}}

\figcaption[fig7.ps]
{Velocity field superposed on \ha monochromatic flux. \label{fig:vit+mono}}

\figcaption[fig8.ps]
{{\bf a)} Optical rotation curve of NGC 5585. \label{fig:rc_Ha}}

\figcaption[fig9.ps]
{{\bf a)} Best fit mass model for NGC 5585 using the \ha rotation curve at
5 \arcsec resolution.
The model parameters are: \mlb = 0.8, $r_c$ = 3.7 kpc and $\sigma$ = 48 \kms. \\
{\bf b)} Dark--to--luminous mass ratio as a function of radius.\label{fig:Ha}}

\figcaption[fig10.ps]
{{\bf a)} Adopted best fit mass model for NGC 5585 using the \ha rotation
curve for r $<$ 120\arcsec\, and the \hI rotation curve for r $>$ 120\arcsec.
The model parameters are: \mlb = 0.8, $r_c$ = 3.9 kpc and $\sigma$ = 53.3 \kms. \\
{\bf b)} Dark--to--luminous mass ratio as a function of radius.\label{fig:HI+Ha}}

\plotone{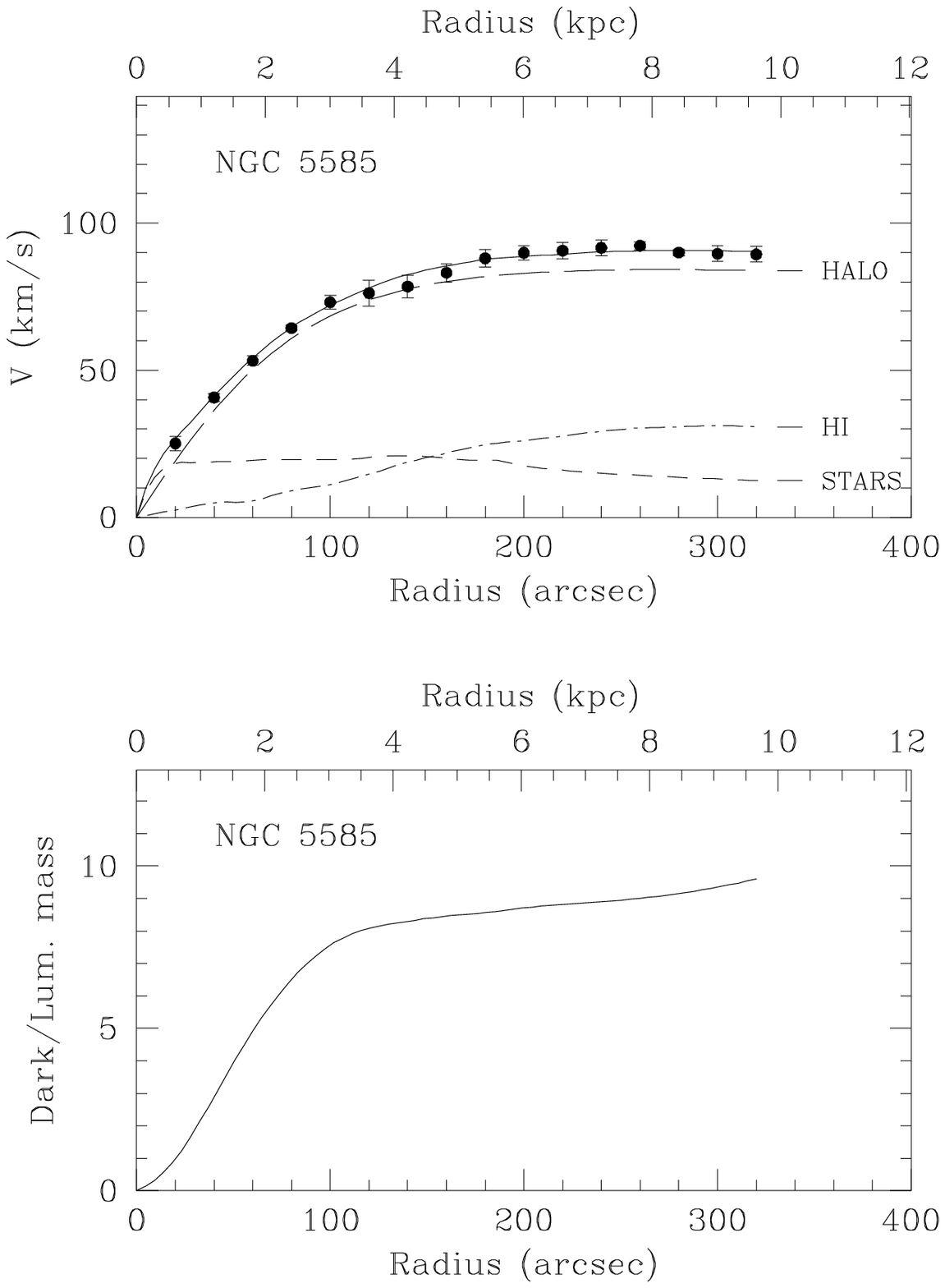}

\plotone{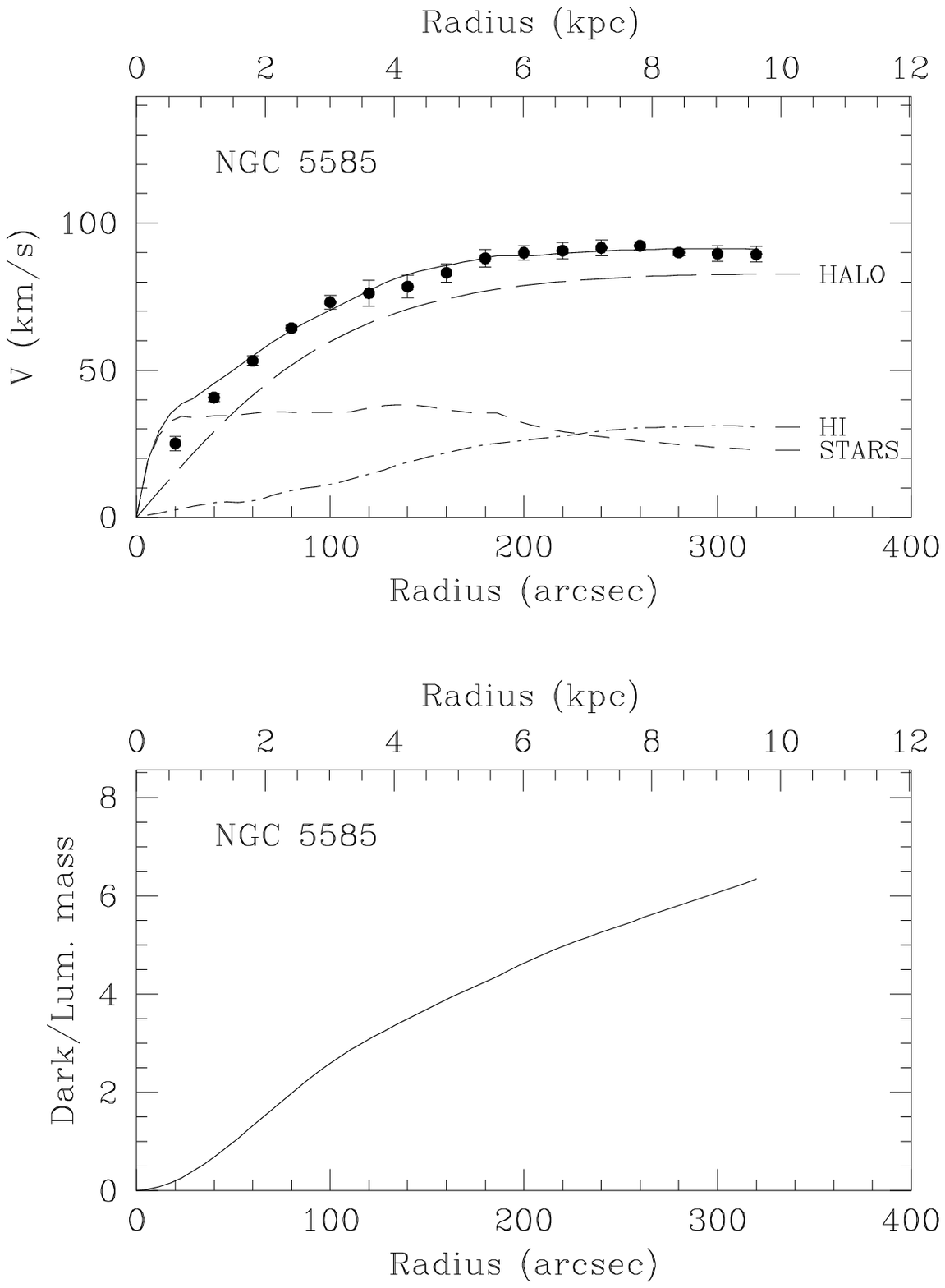}

\plotone{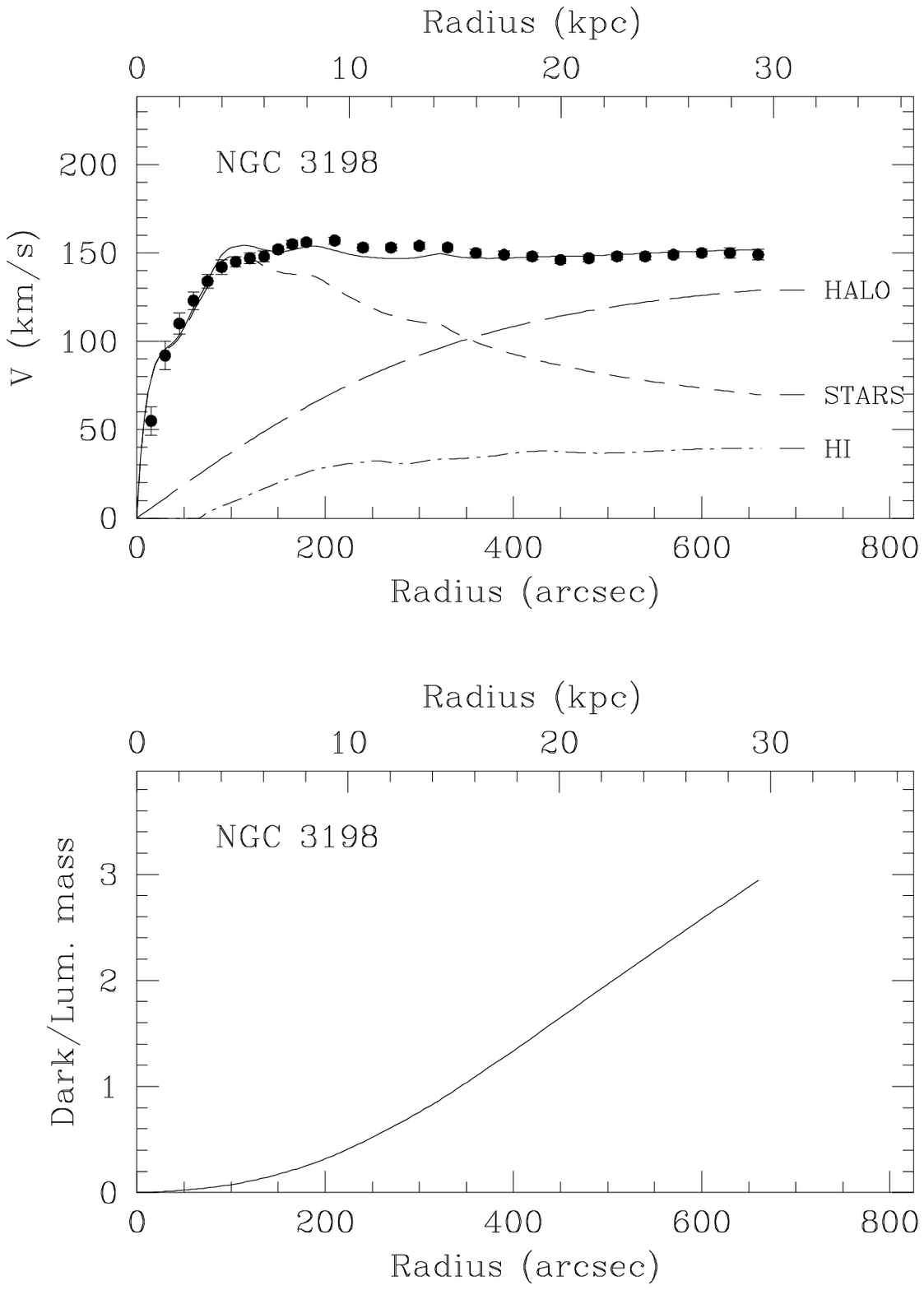}

\plotone{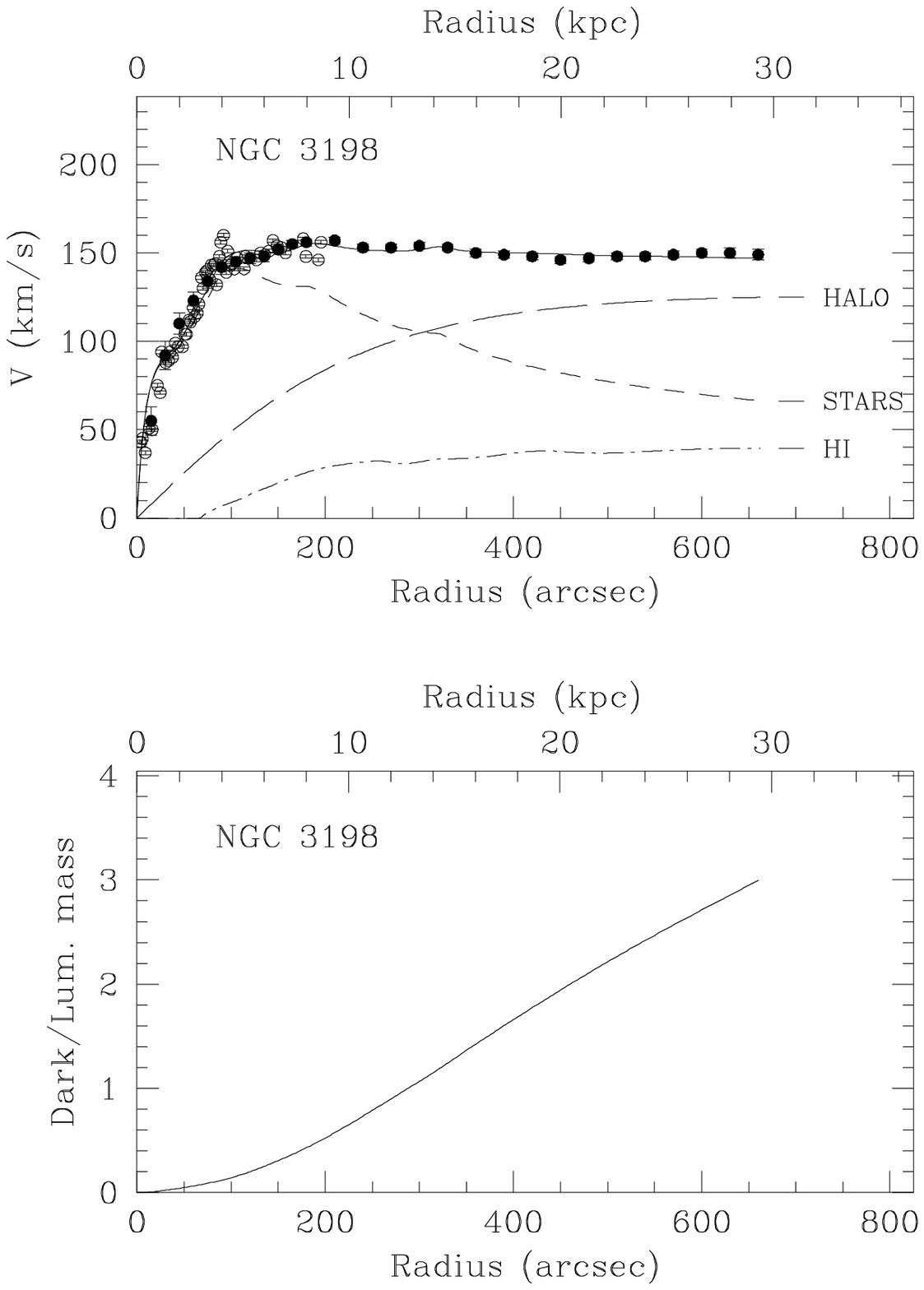}

\plotone{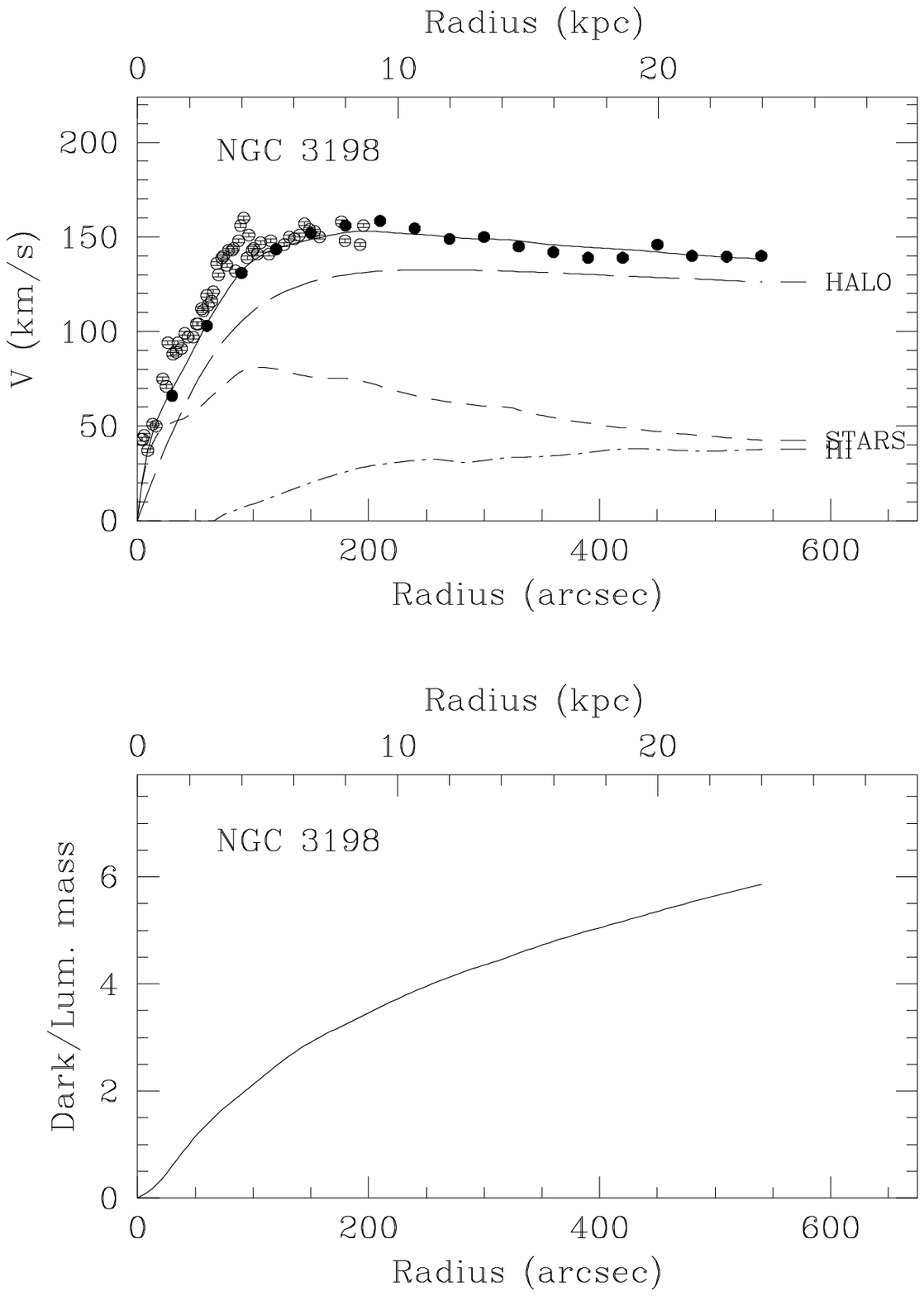}

\plotone{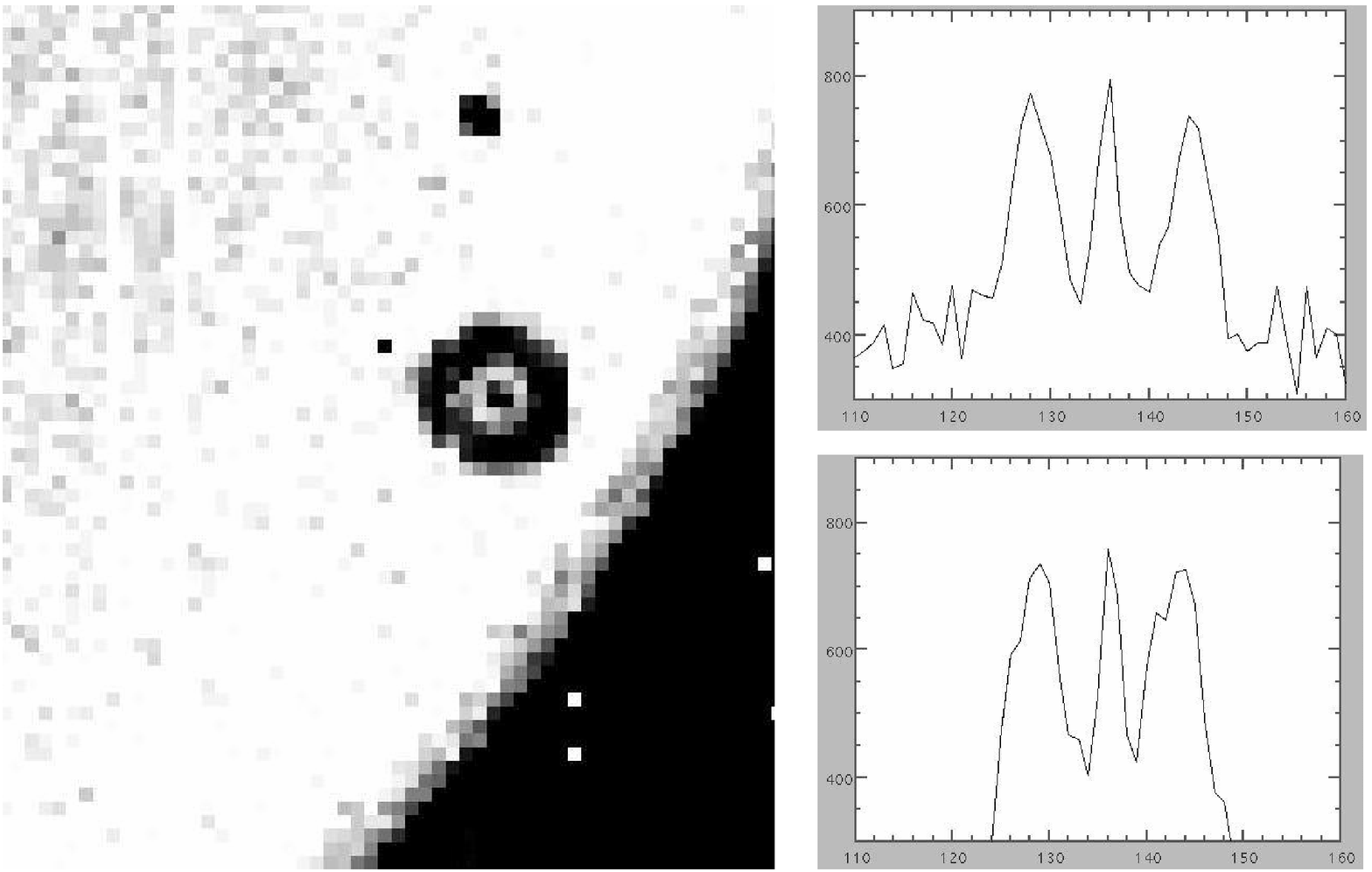}

\plotone{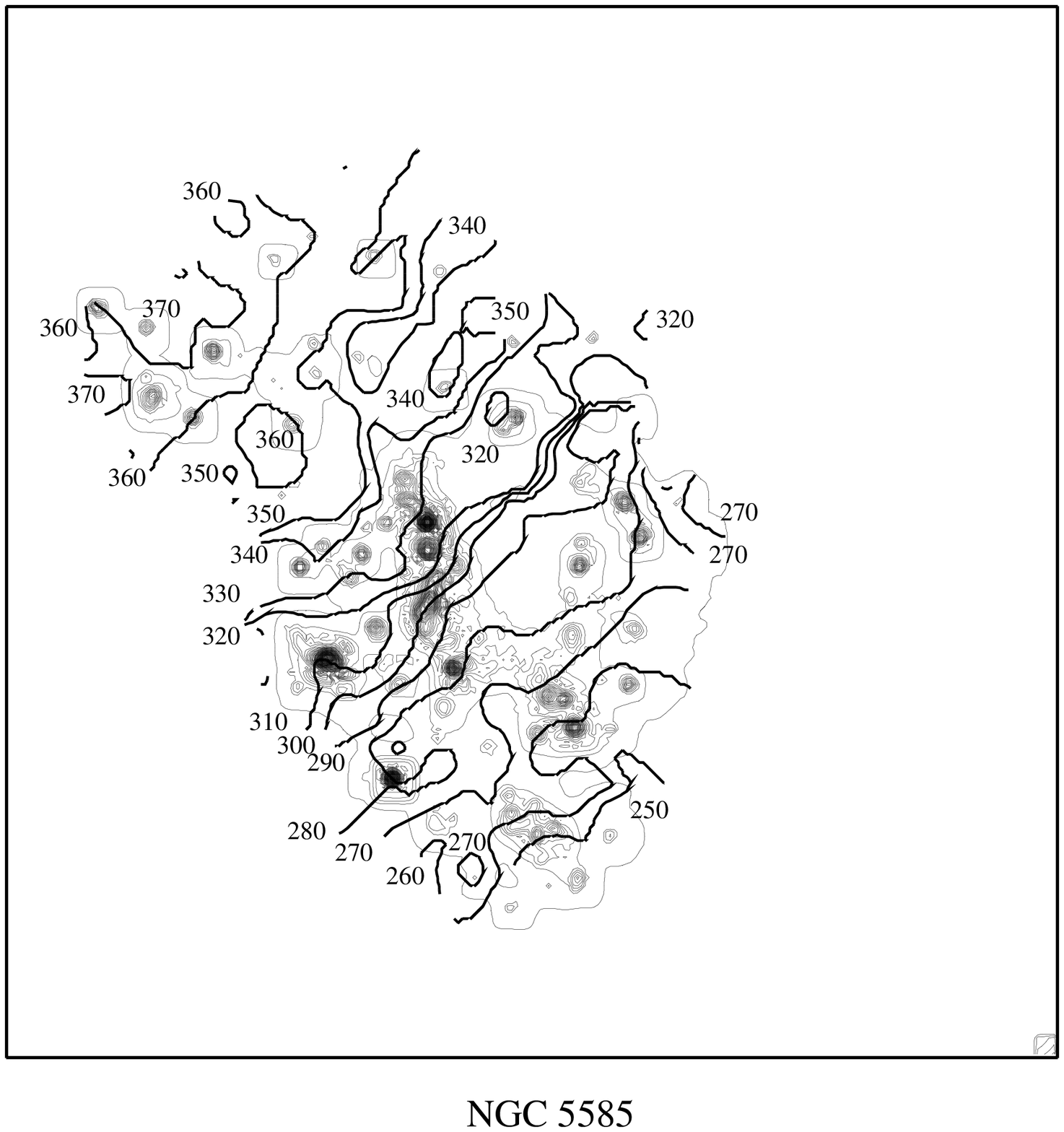}

\plotone{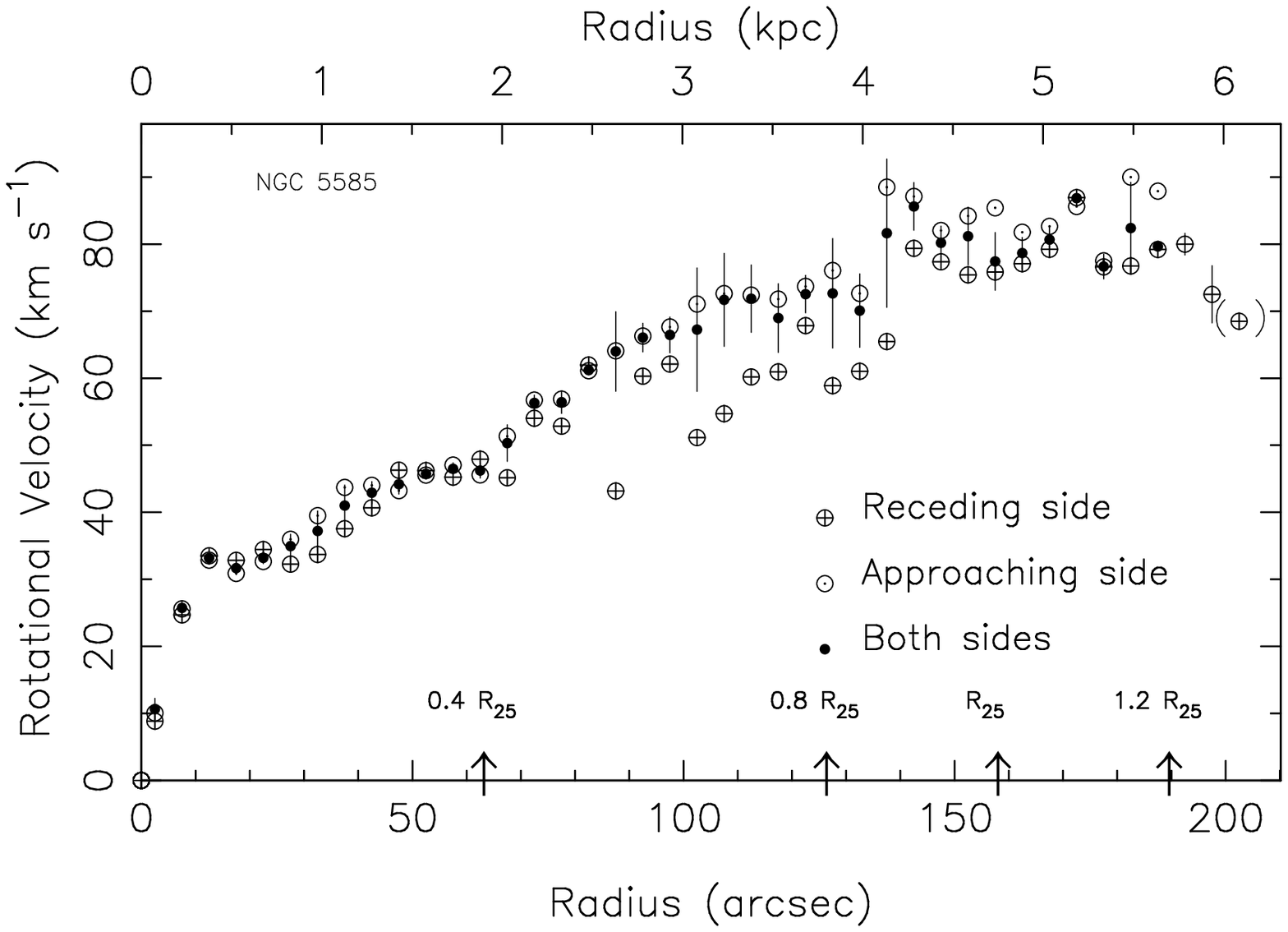}

\plotone{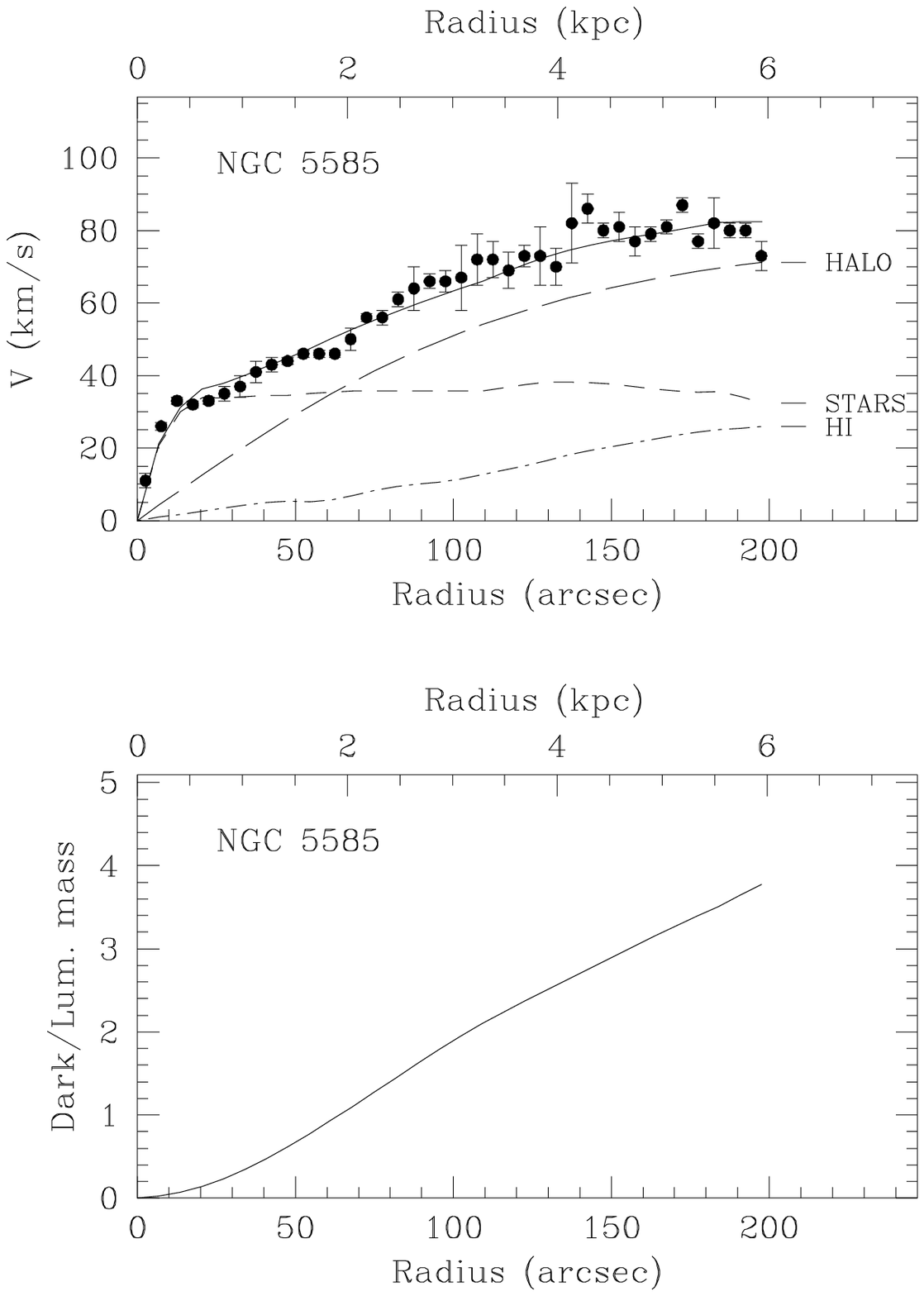}

\plotone{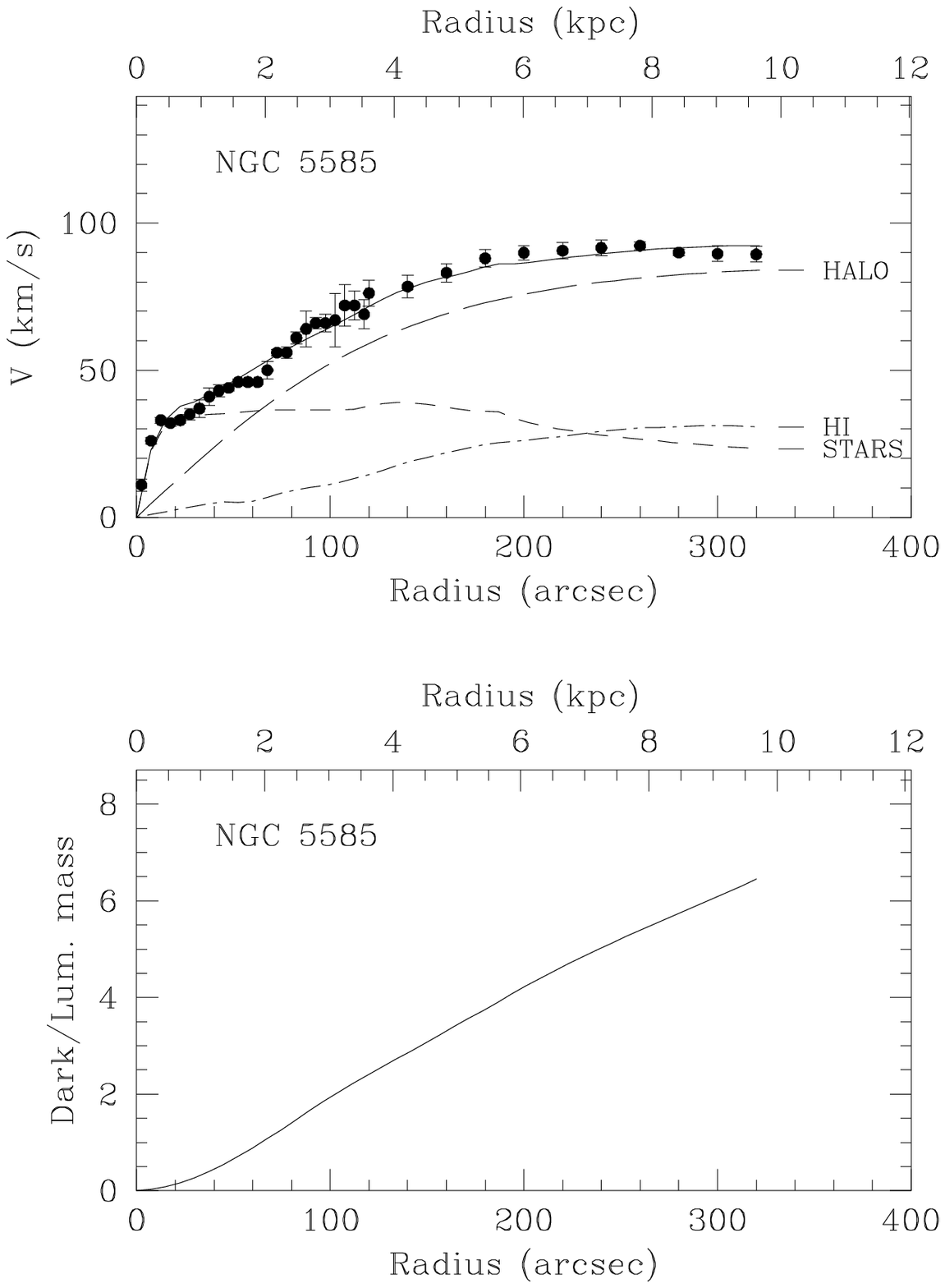}

\end{document}